\documentclass[a4paper,12pt]{article}
\pdfoutput=1
\usepackage{epsfig}
\usepackage{amssymb}
\usepackage{amsfonts}
\usepackage{amsmath}
\usepackage{euscript}
\usepackage{verbatim}
\usepackage{latexsym}
\usepackage{graphicx}

\newskip\humongous \humongous=0pt plus 1000pt minus 1000pt

\newif\ifdtup

\catcode`\@=11

\@addtoreset{equation}{section}

\def\@normalsize{\@setsize\normalsize{15pt}\xiipt\@xiipt
\abovedisplayskip 14pt plus3pt minus3pt%
\belowdisplayskip \abovedisplayskip
\abovedisplayshortskip \z@ plus3pt%
\belowdisplayshortskip 7pt plus3.5pt minus0pt}

\def\small{\@setsize\small{13.6pt}\xipt\@xipt
\abovedisplayskip 13pt plus3pt minus3pt%
\belowdisplayskip \abovedisplayskip
\abovedisplayshortskip \z@ plus3pt%
\belowdisplayshortskip 7pt plus3.5pt minus0pt
\def\@listi{\parsep 4.5pt plus 2pt minus 1pt
     \itemsep \parsep
     \topsep 9pt plus 3pt minus 3pt}}

\relax

\catcode`@=12

\catcode`\@=11

\def\section{\@startsection{section}{1}{\z@}{3.5ex plus 1ex minus
   .2ex}{2.3ex plus .2ex}{\large\bf}}

\def\SymBoxes#1#2#3#4{\newdimen\un@t \un@t#3%
\raisebox{#1}{\rule{#2\un@t}{#4}\hskip-#2\un@t
\@tempdimb\un@t \advance\@tempdimb by-#4\@tempcntb#2\relax%
\@whilenum{\@tempcntb>0}\do{
\rule{#4}{\un@t}\hskip\@tempdimb \advance\@tempcntb by\m@ne}%
\hskip-#2\un@t \rule[\un@t]{#2\un@t}{#4}%
\rule[\un@t]{#4}{#4}\hskip-#4
\rule{#4}{\un@t}}\hskip-#4}                

\begin{document}

\newcommand{\beq}{\begin{equation}}
\newcommand{\eeq}{\end{equation}}
\newcommand{\bea}{\begin{eqnarray}}
\newcommand{\eea}{\end{eqnarray}}
\newcommand{\beas}{\begin{eqnarray*}}
\newcommand{\eeas}{\end{eqnarray*}}
\newcommand{\defi}{\stackrel{\rm def}{=}}
\newcommand{\non}{\nonumber}
\newcommand{\bquo}{\begin{quote}}
\newcommand{\enqu}{\end{quote}}
\renewcommand{\(}{\begin{equation}}
\renewcommand{\)}{\end{equation}}
\def \eqn#1#2{\begin{equation}#2\label{#1}\end{equation}}
\def\IZ{{\mathbb Z}}
\def\IR{{\mathbb R}}
\def\IC{{\mathbb C}}
\def\IQ{{\mathbb Q}}
\def\de{\partial}
\def\Tr{ \hbox{\rm Tr}}
\def\H{ \hbox{\rm H}}
\def\HE{ \hbox{$\rm H^{even}$}}
\def\HO{ \hbox{$\rm H^{odd}$}}
\def\K{ \hbox{\rm K}}
\def\Im{ \hbox{\rm Im}}
\def\Ker{ \hbox{\rm Ker}}
\def\const{\hbox {\rm const.}}
\def\o{\over}
\def\im{\hbox{\rm Im}}
\def\re{\hbox{\rm Re}}
\def\bra{\langle}\def\ket{\rangle}
\def\Arg{\hbox {\rm Arg}}
\def\Re{\hbox {\rm Re}}
\def\Im{\hbox {\rm Im}}
\def\exo{\hbox {\rm exp}}
\def\diag{\hbox{\rm diag}}
\def\longvert{{\rule[-2mm]{0.1mm}{7mm}}\,}
\def\a{\alpha}
\def\dag{{}^{\dagger}}
\def\tq{{\widetilde q}}
\def\p{{}^{\prime}}
\def\W{W}
\def\N{{\cal N}}
\def\hsp{,\hspace{.7cm}}

\def\br{\nonumber\\}
\def\IZ{{\mathbb Z}}
\def\IR{{\mathbb R}}
\def\IC{{\mathbb C}}
\def\IQ{{\mathbb Q}}
\def\IP{{\mathbb P}}
\def \eqn#1#2{\begin{equation}#2\label{#1}\end{equation}}

\newcommand{\C}{\ensuremath{\mathbb C}}
\newcommand{\Z}{\ensuremath{\mathbb Z}}
\newcommand{\R}{\ensuremath{\mathbb R}}
\newcommand{\rp}{\ensuremath{\mathbb {RP}}}
\newcommand{\cp}{\ensuremath{\mathbb {CP}}}
\newcommand{\vac}{\ensuremath{|0\rangle}}
\newcommand{\vact}{\ensuremath{|00\rangle}                    }
\newcommand{\oc}{\ensuremath{\overline{c}}}
\begin{titlepage}
\begin{flushright}
\end{flushright}
\bigskip
\def\thefootnote{\fnsymbol{footnote}}

\begin{center}
{\Large
{\bf A Grassmann Path From AdS$_3$ to Flat Space
}
}
\end{center}

\bigskip
\begin{center}
{\large  Chethan KRISHNAN$^a$\footnote{\texttt{chethan@cts.iisc.ernet.in}}, Avinash RAJU$^a$\footnote{\texttt{avinash@cts.iisc.ernet.in}} and   
Shubho ROY$^a$\footnote{{\texttt{sroy@het.brown.edu}}}
}
\vspace{0.1in}

\end{center}

\renewcommand{\thefootnote}{\arabic{footnote}}

\begin{center}
$^a$ {Center for High Energy Physics\\
Indian Institute of Science, Bangalore, India}\\

\end{center}

\noindent
\begin{center} {\bf Abstract} \end{center}
We show that interpreting the inverse AdS$_3$ radius $1/l$ as a Grassmann variable results in a formal map from gravity in AdS$_3$ to gravity in flat space. 
The underlying reason for this is the fact that $ISO(2,1)$ is the Inonu-Wigner contraction of $SO(2,2)$. We show how this works for the Chern-Simons actions, demonstrate how the general (Banados) solution in AdS$_3$ maps to the general flat space solution, and how the Killing vectors, charges and the Virasoro algebra in the Brown-Henneaux case map to the corresponding quantities in the BMS$_3$ case. 
Our results straightforwardly generalize to the higher spin case: the recently constructed flat space higher spin theories emerge automatically in this approach from their AdS counterparts. We conclude with a discussion of singularity resolution in the BMS gauge as an application.

\vspace{1.6 cm}
\vfill

\end{titlepage}

\setcounter{footnote}{0}

\section{Introduction}

The AdS/CFT correspondence has provided us with substantial insight into the nature of quantum gravity when there is a negative cosmological constant. This includes the possibility of a resolution of the black hole information paradox, and potential exact candidates for quantum gravity in terms of non-gravitational quantum gauge theories. 

Eventually, one would like to understand flat space quantum gravity as well, but taking the vanishing comsological constant limit of the AdS/CFT correspondence in order to accomplish this 
has remained a challenge. 
Some progress in this direction has been made by Barnich and collaborators \cite{Barnich:2010eb, Barnich:2012aw, Barnich:2012xq}\footnote{A very recent work on this topic is \cite{Fareghbal:2013ifa}. See also \cite{Balachandran:2013wsa, Strominger:2013jfa} for some recent interesting thoughts on the asymptotics of flat space.} in the AdS$_3$ case. Specifically, Barnich, Gomberoff and Gonzalez \cite{Barnich:2012aw} showed that the asymptotic symmetry algebra of AdS$_3$ (the Virasoro algebra of Brown-Henneaux) turns into that of flat 2+1 dimensional space (namely, the centrally extended version \cite{Barnich:2006av} of the so-called BMS$_3$ \cite{Ashtekar:1996cd} algebra) in a certain scaling limit where the cosmological constant is sent to zero. 

In this paper, we will show that there is a simple {\em algebraic} way to relate semi-classical gravity in flat space to that in AdS when the spacetime is 2+1 dimensional. The starting point is the fact that 2+1 dimensional gravity can be thought of as a Chern-Simons gauge theory. The gauge group of the theory is $SO(2,2)$ when there is a cosmological constant $\Lambda \equiv -\lambda < 0$, but when $\Lambda = 0$ the gauge group is $ISO(2,1)$. It turns out that an Inonu-Wigner contraction on the  $SO(2,2)$ algebra gives us the $ISO(2,1)$ algebra. This Inonu-Wigner contraction and its connection the BMS/GCA correspondence has been studied in \cite{Bagchi:2010zz, Bagchi:2012cy, Afshar:2013bla}.

Our simple observation is that this Inonu-Wigner contraction of the algebras can be realized at the level of the theories, by taking the inverse AdS$_3$ radius $\epsilon \equiv 1/l =\sqrt{\lambda}$ to be a Grassmann parameter such that $\epsilon^2=0$. 
We show that this trick can be used to map the actions, the solutions and the asymptotic symmetry algebras. Specifically, the general Fefferman-Graham solution for AdS$_3$ gravity written down by Banados goes over into the general flat space solution in BMS gauge, and the Virasoro algebra with the Brown-Henneaux central charge goes over into the BMS$_3$ algebra of flat space with the correct central charge. 

We also show that this approach generalizes to higher spin theories which are essentially Chern-Simons theories with higher rank gauge groups. The recently constructed flat-space higher spin theories emerge very simply and straightforwardly from this approach.  As an illustration of the usefulness of our approach, we show how we can resolve singularities in flat space gravity using higher spins in a BMS-like gauge. We claim that our construction is more ``advantageous" than various other implementations of the limit/contraction, in particular, in the case of higher spin gravity. One reason for this is the fact that our approach can be implemented algebraically. Another (technical) reason is that our approach automatically provides us with a useful trace form in  the Chern-Simons formulation of flat space theory. Since the observables are nonlocal gauge theory objects like holonomies of Wilson loops 
our approach provides an instantly readable/executable map to read them off, unlike in the previous approaches.

\section{Chern-Simons Gravity in Flat Space and AdS}

Witten \cite{Witten:1988hc} noticed that gravity in 2+1 dimensions can be written as a Chern-Simons gauge theory, with gauge group $SO(2,2)$ when the cosmological constant $\Lambda$ is negative and gauge group $ISO(2,1)$ when it is zero. 

We will start with the flat space theory. Our goal is to reproduce the Einstein-Hilbert action in the first order formulation from a Chern-Simons gauge theory. The triad and the spin connection are taken in the form
\bea
e^a=e^a_\mu \ dx^\mu, \ \ \omega^a=\frac{1}{2} \epsilon^{abc} \omega_{\mu bc}\ dx^{\mu}. \label{dualspincon}
\eea
The tangent space indices are raised and lowered using the 2+1 Minkowski metric ${\rm diag}(-1,1,1)$.
Now the claim is that the Chern-Simons action
\bea
I_{CS}[{\cal A}]=\frac{k}{4 \pi}\int {\rm Tr} \left({\cal A} \wedge d {\cal A}+\frac{2}{3}{\cal A} \wedge {\cal A} \wedge {\cal A}\right)
\eea
with 
\bea
{\cal A}\equiv e^a\ P_a +\omega^a\ J_a
\eea
is the Einstein-Hilbert action (with zero cosmological constant) in the first order formulation, if the generators satisfy the $ISO(2,1)$ algebra
\begin{equation}
\left[P_{a},P_{b}\right]=0,\qquad\left[J_{a},J_{b}\right]=\epsilon_{abc}J^{c},\qquad\left[J_{a},P_{b}\right]=\epsilon_{abc}P^{c},\label{eq: ISO(2,1) algebra}
\end{equation}
with $\epsilon^{012}=1$ and the invariant non-degenerate bilinear form
is defined by,
\begin{equation}
{\rm Tr}( J_{a} \ P_{b})=\eta_{ab}, \ \ {\rm Tr}( J_{a} \ J_{b})=0={\rm Tr}( P_{a} \ P_{b}). \label{eq: ISO(2,1) metric}
\end{equation}
Here, the level $k$ of the Chern-Simons theory is related to Newton's constant by
\bea
k=\frac{1}{4G}.
\eea
Once crucial ingredient here worthy of note is the choice of the trace form. For all components of the gauge field to have appropriate kinetic terms, it is necessary that the trace form is non-degenerate. 

Now we turn to gravity with a negative cosmological constant $\Lambda\equiv -\lambda<0$. In this case, Witten's  observation is that again the Einstein-Hilbert action (this time including the cosmological constant piece) can be obtained from the Chern-Simons action and identical definitions as above, if one simply changes the algebra of the $P_a$ and $J_a$ to the $SO(2,2)$ algebra:
\bea
\left[P_{a},P_{b}\right]=\lambda \epsilon_{abc}J^c,\qquad\left[J_{a},J_{b}\right]=\epsilon_{abc}J^{c},\qquad\left[J_{a},P_{b}\right]=\epsilon_{abc}P^{c}.\label{eq: SO(2,2) algebra}
\eea 
In particular, the trace form is the same as before. 

There is a slightly different way of writing this latter (negative cosmological constant) case, that is often used in the literature and we will find convenient. One first introduces the generators
\bea
J_a^{\pm}=\frac{1}{2}\left(J^a\pm l\ P^a \right), \label{pmJ}
\eea
where $l=\frac{1}{\sqrt{\lambda}}$.
It is easy to check that (\ref{eq: SO(2,2) algebra}) now takes the form
\bea
\left[J_{a}^+,J_{b}^-\right]=0,\qquad\left[J_{a}^+,J_{b}^+\right]=\epsilon_{abc}J^{c+},\qquad\left[J_{a}^-,J_{b}^-\right]=\epsilon_{abc}J^{c-}. \label{sl2 factorized form}
\eea
The first of the above commutators implies that the algebra is a direct sum: what we have essentially shown is that $SO(2,2)\sim SL(2,\IR)\times SL(2,\IR)$, and that its algebra can be written as a direct sum of two copies of ${\bf sl}(2,\IR)$. In particular this means that we can introduce $T^a$ and $\tilde T^a$ via
\bea
J_{a}^+=\left(\begin{array}{cc}
T^a&0\\
 0 & 0
\end{array}\right),\qquad J_{a}^-=\left(\begin{array}{cc}
0&0\\
 0 & \tilde T^a
\end{array}\right)
\eea
so that if $T^a$ and $\tilde T^a$ each satisfy the $SL(2,\IR)$ algebra,
\bea
\left[T_{a},T_{b}\right]=\epsilon_{abc}T^{c},\qquad\left[\tilde T_{a},\tilde T_{b}\right]=\epsilon_{abc}\tilde T^{c}. \label{T-tilde-T form}
\eea
then (\ref{sl2 factorized form}), and therefore (\ref{eq: SO(2,2) algebra}), are satisfied. An important point to note is that from the trace form (\ref{eq: ISO(2,1) metric}) one finds that the trace form in terms of  $T$ and $\tilde T$ are
\bea
{\rm Tr}(T_a T_b)=\frac{l}{2} \eta_{ab}, \ \ \tilde {\rm Tr}(\tilde T_a \tilde T_b)=-\frac{l}{2} \eta_{ab}, 
\eea
In terms of $T$ and $\tilde T$, the gauge field now takes the form
\bea
{\cal A}_\mu=\left(\begin{array}{cc}
\left(\omega_\mu^a+\frac{1}{l}e_\mu^a\right)\ T_a&0\\
 0 & \left(\omega_\mu^a-\frac{1}{l}e_\mu^a\right)\ \tilde T_a
\end{array}\right)\equiv \left(\begin{array}{cc}
A_\mu^a T_a&0\\
 0 & \tilde A_\mu^a \tilde T_a
\end{array}\right)
\eea
so that the Einstein-Hilbert action with a cosmological constant can be written as the sum of two pieces now:
\bea
\frac{k}{4 \pi}\int {\rm Tr} \left({ A} \wedge d { A}+\frac{2}{3}{ A} \wedge { A} \wedge { A}\right)+\frac{k}{4 \pi}\int \tilde{\rm Tr} \left({ \tilde A} \wedge d {\tilde A}+\frac{2}{3}{\tilde A} \wedge {\tilde A} \wedge {\tilde A}\right)
\eea

Since the algebra of both $T$'s and $\tilde T$'s is identical (namely the $SL(2,\IR)$ algebra), what is typical in the literature is to identify the generator matrices $T^a = \tilde T^a$. This means that their trace forms are also identical, which one takes to be
\bea
{\rm Tr}(T_a T_b)=\frac{1}{2} \eta_{ab}. \label{simpltr}
\eea
Note that this trace form does not have the factor of $l$ as before, so that the missing $l$ has to be incorporated into the Chern-Simons level by hand for the action to reduce to the Einstein-Hilbert form. So now
\bea
k=\frac{l}{4G}.
\eea
Also, the negative sign in the trace form of the $\tilde T$ should also be incoprorated into the action by hand, so that now the AdS Einstein-Hilbert action takes the final form
\bea
I_{EH_{AdS}}=\frac{k}{4 \pi}\int {\rm Tr} \left({ A} \wedge d { A}+\frac{2}{3}{ A} \wedge { A} \wedge { A}\right)-\frac{k}{4 \pi}\int {\rm Tr} \left({ \tilde A} \wedge d {\tilde A}+\frac{2}{3}{\tilde A} \wedge {\tilde A} \wedge {\tilde A}\right) \label{adsaction}
\eea 
where now the $A$ and $\tilde A$ are understood to be expanded in a basis of $T^a$'s (and no $\tilde T^a$'s):
\bea
A_\mu=\left(\omega_\mu^a+\frac{1}{l}e_\mu^a\right)\ T_a, \ \ \tilde A_\mu=\left(\omega_\mu^a-\frac{1}{l}e_\mu^a\right)\ T_a \label{gaugef}
\eea
with trace form (\ref{simpltr}).


The basic reason why we have set up these constructions carefully is because the precise chain of logic in writing down the action in the form (\ref{adsaction}) is often not discussed in the literature, but is crucial for what we are about to discuss. One basic observation in this paper is that if one makes the replacement 
\bea
\frac{1}{l} \rightarrow \epsilon
\eea
(where $\epsilon$ is a Grassmann parameter so that 
$\epsilon^2=0$),
in (\ref{gaugef}), then the AdS Einstein-Hilbert action (\ref{adsaction}) turns into the flat space Einstein-Hilbert action, but multiplied by an overall factor of $\epsilon$. In other words we will see that the quantity multiplying the $\epsilon$, after the above replacement, is the flat space gravitational action. This makes sure that the Newton's constant and Chern-Simons level after this replacement are related by
\bea
k =\frac{1}{4G}. \label{grassG}
\eea
Even though we will not do so here, we can absorb the overall factor of $\epsilon$ into the definition of the $G$ and formally treat $G$ as a Grassmann parameter: since we are mostly interested in classical equations of motion where $G$ is merely an overall factor, this will not make any difference at the level of the solutions.  
These claims are easy to check by direct computation, and we have done so. 

For most purposes we will be using this map from AdS to flat space as a useful technical tool for dealing with various aspects of classical solutions, so for the purposes of this paper, we will think of it as a formal tool. But the simplifications that happen are sufficiently drastic, that it is tempting to speculate that there is more to this story than a mere trick. 



The fundamnetal reason why the above replacement works is because of the fact that $ISO(2,1)$ is an Inonu-Wigner contraction of $SO(2,2)$. For the specific case here, Inonu-Wigner contraction is the statement that if one scales the generators $P^a$ in the $SO(2,2)$ algebra (\ref{eq: SO(2,2) algebra}) by a (non-Grassmann) parameter $\epsilon$ (that is $P^a \rightarrow \epsilon P^a$) and then takes $\epsilon \rightarrow 0$, one is left with the $ISO(2,1)$ algebra (\ref{eq: ISO(2,1) algebra}). But instead of taking the {\em analytic} limit $\epsilon \rightarrow 0$ to implement the Inonu-Wigner contraction, one can also treat $\epsilon$ as a Grassmann parameter and end with the same (\ref{eq: SO(2,2) algebra}). This is an {\em algebraic} realization of the contraction and that is what we are putting to use here. 

An explicit way in which both the norms and the algebras of $ISO(2,1)$ can be realized in terms of the $T^a$ and $\tilde T^a$ generators of $SL(2,\IR)$ is to define:
\bea
P^a=\left(\begin{array}{cc}
\epsilon \ T^a&0\\
 0 & -\epsilon \ \tilde T^a
\end{array}\right), \ &&\ J^a=\left(\begin{array}{cc}
 T^a&0\\
 0 &  \tilde T^a
\end{array}\right)
\eea
If one identifies $T^a$ with $J^{a+}=\Big(\begin{array}{cc}
T^a&0\\
 0 & 0
\end{array}\Big)$ and $\tilde T^a$ with $J^{a-}=\Big(\begin{array}{cc}
0&0\\
 0 & \tilde T^a
\end{array}\Big)$, then this can be thought of as another way to write
\bea
P^a=\epsilon(T^a-\tilde T^a), \ \ J^a=(T^a+\tilde T^a)
\eea
which in turn follows from (\ref{pmJ}) upon $1/l \rightarrow \epsilon$. This generalizes very straightforwardly to higher spin theories as well, as we will briefly discuss later.

Another (non-Grassmann) way to think of the mapping from one theory to other is to think of it as the scaling limit where $1/l \rightarrow 0$ but with $k/l$ is held fixed. Even though it is not couched there in this language, this is essentially what BGG have done \cite{Barnich:2012aw}. We will find this useful in our discussion of the Brown-Henneaux algebra.


It is trivial to check that the diffeomorphisms and local Lorentz transformations, written in terms of the triads and the spin connections, also go over from the AdS to the flat case without any difficulty when we set $1/l \rightarrow \epsilon$. The explicit expressions can be found in Witten's paper and the check is trivial, so we will not repeat them here.

Often, in what follows we will use generators $T^a$ that have the trace form
\bea
{\rm Tr}(T^aT^b)=2 \eta^{ab},
\eea
following the conventions of \cite{Castro:2011fm}, where they are working with 3 $\times$ 3 generators, which are more convenient from the perspective of generalizations to higher spin theories. 
This implies that we should take
\bea
k=\frac{l}{16 G},
\eea
in the AdS case.

\section{AdS$_{3}$ in BMS-like gauge}

Our goal is first to show the transition from general locally AdS$_3$ 
solution \cite{Banados:1998gg} to the general asymptotically flat solution using the Grassmann approach\footnote{In this section, we have chosen to set $8G=1$ in agreement with the general convention
in $2+1$-d general relativity literature.%
}. 

Following \cite{Barnich:2012aw}, we first write down the general locally AdS solution in a BMS-like gauge to ease the transition to flat space. The general asymptotically AdS$_3$ line element that satisfies the Einstein equations with a negative cosmological constant can be written in the form
\begin{equation}
ds^{2}=\left(\mathcal{M}-\frac{r^{2}}{l^{2}}\right)du^{2}-2dudr+2\mathcal{N}\: dud\phi+r^{2}d\phi^{2}\label{eq: AdS_3 metric in BMS gauge}
\end{equation}
provided 
\begin{equation}
\partial_{u}\mathcal{M}=\frac{2}{l^{2}}\partial_{\phi}\mathcal{N},\qquad2\partial_{u}\mathcal{N}=\partial_{\phi}\mathcal{M}.\label{eq: neccessary conditions on M,N}
\end{equation}
This solution, and these conditions on the arbitrary functions are merely a re-writing of the general Fefferman-Graham solution in AdS$_3$ \cite{Banados:1998gg}. 
This is easily checked by noting that $\mathcal{M},\mathcal{N}\equiv \mathcal{M}(\phi,u),\mathcal{N}(\phi,u)$ satisfying the above conditions
can be expressed in terms of the usual left and right moving functions,
$\mathcal{L}(x^{+}),\bar{\mathcal{L}}(x^{-})$, 
\begin{equation}
\mathcal{M}(u,\phi)=2\left(\mathcal{L}(x^{+})+\bar{\mathcal{L}}(x^{-})\right), \ \ \mathcal{N}(u,\phi)=l\left(\mathcal{L}(x^{+})-\bar{\mathcal{L}}(x^{-})\right),\label{eq: BMS gauge AdS parameter}
\end{equation}
with $x^{\pm}=\frac{u}{l}\pm\phi$. 

We take the triad for this locally AdS$_{3}$ solution (in BMS like coordinates) to be%
, 
\bea
e=-\frac{1}{\sqrt{2}}\left[\left(\frac{\mathcal{M}}{2}-1-\frac{r^{2}}{2l^{2}}\right)\: du-dr+\mathcal{N}\: d\phi\right]T_{0}+\hspace{1in}\nonumber \\ \hspace{1in}-\frac{1}{\sqrt{2}}\left[\left(\frac{\mathcal{M}}{2}+1-\frac{r^{2}}{2l^{2}}\right)\: du-dr+\mathcal{N}\: d\phi\right]T_{1}
-rd\phi T_{2}.\label{eq: BMS gauge AdS triad}
\eea
The (dualized) spin-connection (\ref{dualspincon}) can be computed directly from the triads, and the result is:%
\begin{eqnarray*}
\omega^{0} & = & -\frac{1}{\sqrt{2}}\left(\frac{\partial_{u}\mathcal{N}}{r}-\frac{\partial_{\phi}\mathcal{M}}{2r}+\frac{\mathcal{N}}{l^{2}}\right)du-\frac{1}{\sqrt{2}}\left(\frac{\mathcal{M}}{2}-1-\frac{r^{2}}{2l^{2}}\right)d\phi,\\
\omega^{1} & = & -\frac{1}{\sqrt{2}}\left(\frac{\partial_{u}\mathcal{N}}{r}-\frac{\partial_{\phi}\mathcal{M}}{2r}+\frac{\mathcal{N}}{l^{2}}\right)du-\frac{1}{\sqrt{2}}\left(\frac{\mathcal{M}}{2}+1-\frac{r^{2}}{2l^{2}}\right)d\phi,\\
\omega^{2} & = & -\frac{r}{l^{2}}du.
\end{eqnarray*}
These expressions have been checked by hand to satisfy the torsion-free
condition,
\[
\partial_{\mu}e_{\nu}^{a}-\partial_{\nu}e_{\mu}^{a}+\epsilon^{a}\,_{bc}\left(e^{b}\,_{\mu}\omega^{c}\,_{\nu}-e^{b}\,_{\nu}\omega^{c}\,_{\mu}\right)=0,
\]
and the Einstein equation \cite{Witten:1988hc} (provided conditions
(\ref{eq: neccessary conditions on M,N}) hold),
\[
\partial_{\mu}\omega_{\nu}^{a}-\partial_{\nu}\omega_{\mu}^{a}+\epsilon^{a}\,_{bc}\left(\omega^{b}\,_{\mu}\omega^{c}\,_{\nu}+\frac{1}{l^{2}}e^{b}\,_{\mu}e_{\nu}^{c}\right)=0.
\]
Note that the asymptotic AdS$_3$ fall off conditions went into the construction of the Fefferman-Graham form: they are implicit in our starting point. So the constraints (\ref{eq: neccessary conditions on M,N}) came purely from imposing the AdS$_3$ Einstein equations.

Using the triads and the spin connection, now we can immediately write down the explicit gauge field corresponding to the general asymptotically AdS$_3$ solution via (\ref{gaugef}).


\section{Grassmann Path to Flat Space}

Now we turn to the general locally flat solution. In the ``BMS-gauge'' \cite{Barnich:2010eb}, where asymptotic analysis
is easiest (akin to Fefferman-Graham gauge in the case of $AdS$),
the most general solution in $2+1$-d is,
\begin{equation}
ds^{2}=\mathcal{M}(\phi)du^{2}-2dudr+2\left[\mathcal{J}(\phi)+\frac{u}{2}\partial_{\phi}\mathcal{M}(\phi)\right]dud\phi+r^{2}d\phi^{2}.\label{eq: most general asymptotically flat}
\end{equation}
(Later we will specialize to the case when $\mathcal{M}(\phi)=M$
and $\mathcal{J}(\phi)=J/2$ are constants, which has a cosmological interpretation).

Now, the gauge field from the last section, upon the Grassmann replacement of $1/l \rightarrow \epsilon$ gives us the explicit form
\begin{eqnarray}
A & = & -\frac{1}{\sqrt{2}}\left[\epsilon\left(\frac{\mathcal{M}}{2}-1\right)\: du-\epsilon dr+\epsilon\left(\mathcal{J}+\frac{u}{2}\mathcal{M}'\right)\: d\phi+\left(\frac{\mathcal{M}}{2}-1\right)\: d\phi\right]T_{0}\nonumber \\
 &  & \qquad-\frac{1}{\sqrt{2}}\left[\epsilon\left(\frac{\mathcal{M}}{2}+1\right)\: du-\epsilon dr+\epsilon\left(\mathcal{J}+\frac{u}{2}\mathcal{M}'\right)\: d\phi+\left(\frac{\mathcal{M}}{2}+1\right)\: d\phi\right]T_{1}\nonumber \\
 &  & \qquad\qquad\qquad\qquad\qquad\qquad\qquad\qquad\qquad\qquad\qquad\qquad\qquad\qquad\qquad-\epsilon\: r\: d\phi\; T_{2}.\hspace{0.5in}\label{eq: BMS gauge full connection-1}
\end{eqnarray}
Note that the Grassmann replacement gives a simple interpretation for the form of the functions now because of the constraints (\ref{eq: neccessary conditions on M,N}):
\bea
\mathcal{M} \equiv \mathcal{M}(\phi), \ \ \mathcal{N} \equiv \mathcal{J}(\phi)+\frac{u}{2}\mathcal{M}'(\phi).
\eea
Our claim from section 2 is that the $ISO(2,1)$ theory can be reformulated as a Grassmann valued $SO(2,2)$
gauge theory with the connection\footnote{We will exclusively work with ``holomorphic" part. The ``anti-holomorphic" $\tilde A=\left(\omega^{a}-\epsilon e^{a}\right)T_{a}$ part is entirely analogous.}
\[
A=\left(\omega^{a}+\epsilon e^{a}\right)T_{a}
\]
where, $T_{a}\in SO(2,1)=SL(2,R)$ and $\epsilon$ is a Grassman parameter.
This means that we can read off the flat space triad and spin connection from this gauge field. 

Indeed, it is easy to check that the triad and spin connection one obtains this way, can reproduce the most general flat metric  (\ref{eq: most general asymptotically flat})! The natural expressions for the flat space triads are \cite{Gonzalez:2013oaa}%
\footnote{However in contrast to the convention of \cite{Gonzalez:2013oaa},
we choose a convention where $
\eta_{ab}=\mbox{diag}\left(-1,1,1\right)$.
},
\begin{eqnarray}
e^{0} & = & -\frac{1}{\sqrt{2}}\left[\left(\frac{\mathcal{M}(\phi)}{2}-1\right)\: du-dr+\left(\mathcal{J}(\phi)+\frac{u}{2}\frac{d\mathcal{M}(\phi)}{d\phi}\right)\: d\phi\right],\nonumber \\
e^{1} & = & -\frac{1}{\sqrt{2}}\left[\left(\frac{\mathcal{M}(\phi)}{2}+1\right)\: du-dr+\left(\mathcal{J}(\phi)+\frac{u}{2}\frac{d\mathcal{M}(\phi)}{d\phi}\right)\: d\phi\right],\nonumber \\
e^{2} & = & -r\: d\phi,\label{eq: BMS gauge triad}
\end{eqnarray}
We can compute the spin-connection from Cartan's torsion-free condition and this also matches the result obtained from Grassmann replacement from AdS:
\begin{eqnarray}
\omega^{0} & = & -\frac{1}{\sqrt{2}}\left(\frac{\mathcal{M}(\phi)}{2}-1\right)\: d\phi,\nonumber \\
\omega^{1} & = & -\frac{1}{\sqrt{2}}\left(\frac{\mathcal{M}(\phi)}{2}+1\right)\: d\phi,\nonumber \\
\omega^{2} & = & 0.\label{eq: BMS gauge spin-connection}
\end{eqnarray}

For later use we write down the $ISO(2,1)$ connection in terms of the $ISO(2,1)$ generators as well (\ref{eq: ISO(2,1) algebra}):%
\begin{eqnarray}
A & = & -\frac{1}{\sqrt{2}}\left[\left(\frac{\mathcal{M}}{2}-1\right)\: du-dr+\left(\mathcal{J}+\frac{u}{2}\mathcal{M}'\right)\: d\phi\right]P_{0}\nonumber \\
 &  & \qquad\qquad\qquad-\frac{1}{\sqrt{2}}\left[\left(\frac{\mathcal{M}}{2}+1\right)\: du-dr+\left(\mathcal{J}+\frac{u}{2}\mathcal{M}'\right)\: d\phi\right]P_{1}\nonumber \\
 &  & \qquad\qquad\qquad\qquad-r\: d\phi\; P_{2}-\frac{1}{\sqrt{2}}\left(\frac{\mathcal{M}}{2}-1\right)\: d\phi\: J_{0}-\frac{1}{\sqrt{2}}\left(\frac{\mathcal{M}}{2}+1\right)\: d\phi\: J_{1},\label{eq: BMS gauge full connection}
\end{eqnarray}
where, $\mathcal{M}'=\partial_{\phi}\mathcal{M}(\phi)$.


\section{Asymptotic Charge Algebra}

We restore all factors of $8G$ for this section to facilitate a consistent
derivation of the flat space from a grassmanian AdS expressions. The
AdS charges (as derived in BMS looking gauge) were written down by
\cite{Barnich:2012aw}, 
\begin{equation}
Q_{f,Y}=\frac{1}{16\pi G}\int_{0}^{2\pi}d\phi\left[f\left(\mathcal{M}+1\right)+2Y\mathcal{N}\right],\label{eq: AdS charges}
\end{equation}
associated with the killing vector,
\[
\xi_{f,Y}=f\: du-Y-l\partial_{\phi}f\left(\int_{r}^{\infty}dr'r'^{-2}e^{2\beta}\right)\: d\phi-r\left(\partial_{\phi}\xi^{\phi}-U\partial_{\phi}f\right)
\]
with $f,Y$ defined in terms of purely holomorphic and antiholomorphic
(arbitrary) functions, $Y^{\pm}$: 
\[
f=\frac{l}{2}\left(Y^{+}(x^{+})+Y^{-}(x^{-})\right),Y=\frac{1}{2}\left(Y^{+}(x^{+})-Y^{-}(x^{-})\right).
\]
$U,\beta,V$ are metric paramaters,
\[
ds^{2}=e^{2\beta}\frac{V}{r}du^{2}-2e^{2\beta}dudr+r^{2}\left(d\phi-Udu\right)^{2}.
\]
In fact looking at the metric (\ref{eq: AdS_3 metric in BMS gauge}),
we have,
\[
\beta=0,\frac{V}{r}+r^{2}U^{2}=-\frac{r^{2}}{l^{2}}+\mathcal{M},\mathcal{N}=-r^{2}U.
\]
First we do a mode decomposition \cite{Barnich:2012aw},
\[
\mathcal{L}(\bar{\mathcal{L}})=-\frac{1}{4}+\sum_{m}\frac{1}{2l}L^{\pm}e^{-imx^{\pm}},
\]
we have,
\[
\mathcal{M}=-1+\sum_{m}\frac{8G}{l}\left(L_{m}^{+}e^{-imu/l}+L_{-m}^{-}e^{imu/l}\right)e^{-im\phi},
\]
\[
\mathcal{N}=4G\sum_{m}\left(L_{m}^{+}e^{-imu/l}-L_{-m}^{-}e^{imu/l}\right)e^{-im\phi}.
\]
So replacing,
\[
\frac{1}{l}\rightarrow\epsilon,
\]
these mode expansions, become
\begin{eqnarray}
\mathcal{M} & = & -1+8G\epsilon\left(L_{0}^{+}+L_{0}^{-}\right)+8G\epsilon\sum_{m\neq0}\left(L_{m}^{+}(1-\epsilon\: imu)+L_{-m}^{-}(1+\epsilon\: imu)\right)e^{-im\phi}\nonumber \\
 & = & -1+8G\epsilon\sum_{m}\left(L_{m}^{+}+L_{-m}^{-}\right)e^{-im\phi},\nonumber \\
\mathcal{N} & = & 4G\left(L_{0}^{+}-L_{0}^{-}\right)+4G\sum_{m\neq0}\left[\left(L_{m}^{+}-L_{-m}^{-}\right)-\epsilon u\: im\left(L_{m}^{+}+L_{-m}^{-}\right)\right]e^{-im\phi}.\label{eq:Intermediate stage M, N}
\end{eqnarray}
Next we define,
\begin{equation}
P_{m}=\frac{1}{l}\left(L_{m}^{+}+L_{-m}^{-}\right),J_{m}=L_{m}^{+}-L_{-m}^{-},\label{eq: New ISO generators}
\end{equation}
which after the Grassman replacement turns into
 \begin{equation}
\mathcal{P}_{m}=\epsilon\left(L_{m}^{+}+L_{-m}^{-}\right),\mathcal{J}_{m}=L_{m}^{+}-L_{-m}^{-},\label{eq: New ISO generators}
\end{equation}
These definitions can be motivated in two ways. One is by taking a cue from \cite{Barnich:2012xq} and making the replacement $1/l \rightarrow \epsilon$ in the expressions there. Another way to motivate this definition is as follows. These modes are to be thought of as capturing the infinite dimensional extension of $SL(2, \IR)$ (or $ISO(2,1)$ after the flat space limit). The zero mode part of these generators is $SL(2, \IR)$ (respectively $ISO(2,1)$). The definitions, restricted to the zero mode sector is precisely what is needed to make the transition from $SL(2, \IR)$  to $ISO(2,1)$, so it is natural extend the definitions to the higher modes as well. Either way, ultimately the only thing that matters is that this definition ends up giving us BMS$_3$ from Virasoro as we show presently.

With the above definitions,
\begin{equation}
\mathcal{M}(\phi)=-1+8G\sum_{m}P_{m}e^{-im\phi},\label{eq: familiar BMS flat M}
\end{equation}
\begin{equation}
\mathcal{N}=8G\left(\mathcal{J}(\phi)+\frac{u}{2}\partial_{\phi}\mathcal{M}(\phi)\right),\mathcal{J}(\phi)\equiv\frac{1}{2}\sum_{m}J_{m}e^{-im\phi}.\label{eq:familiar BMS flat N}
\end{equation}
Similarly for the killing vector parameters after making the replacements,
\begin{eqnarray*}
\epsilon\: f & = & \frac{1}{2}\sum_{m}\left(Y_{m}^{+}+Y_{-m}^{-}\right)e^{-im\phi}-\epsilon\: u\sum_{m\neq0}im\left(Y_{m}^{+}-Y_{-m}^{-}\right)e^{-im\phi},\\
Y & = & \frac{1}{2}\sum_{m}\left(Y_{m}^{+}-Y_{-m}^{-}\right)e^{-im\phi}-\epsilon\: u\sum_{m\neq0}im\left(Y_{m}^{+}+Y_{-m}^{-}\right)e^{-im\phi}
\end{eqnarray*}
Analogous to (\ref{eq: New ISO generators}) we have
\begin{equation}
\epsilon T_{m}\equiv\frac{1}{2}\left(Y_{m}^{+}+Y_{-m}^{-}\right),Y_{m}\equiv\frac{1}{2}\left(Y_{m}^{+}-Y_{-m}^{-}\right).\label{eq: Definition for Killing modes to go from SL2 to ISO}
\end{equation}
This leads to the expressions,
\begin{equation}
f=T(\phi)+u\partial_{\phi}Y(\phi),Y=Y(\phi)\label{eq: Killing vector parameters for ISO theory}
\end{equation}
Now finally we can plug equations (\ref{eq: familiar BMS flat M}),
(\ref{eq:familiar BMS flat N}), and (\ref{eq: Killing vector parameters for ISO theory})
in the expression (\ref{eq: AdS charges}) for the AdS charges to
obtain,
\begin{equation}
Q_{T,Y}=\frac{1}{16\pi G}\int_{0}^{2\pi}\left(T(\phi)\:\mathcal{M}(\phi)+2Y(\phi)\mathcal{J}(\phi)\right).\label{eq: BMS charges}
\end{equation}
This is exactly the expression of $ISO$ charges obtained in \cite{Barnich:2010eb}
upon conducting a Henneaux-Teitelboim like asymptotic symmetry analysis
for flat space (BMS/CFT correspondence).

To conlude this section we show how the Virasoro algebra with Brown-Henneaux central charge goes over to the BMS algebra with the correct central charge\footnote{See \cite{2005:math} for a related discussion in a different context.}. 
The latter has central charges $c^{\pm}=\frac{3l}{2G}$:
\begin{equation}
\left[L_{m}^{\pm},L_{n}^{\pm}\right]=\left(m-n\right)L_{m+n}^{\pm}+\frac{c^{\pm}}{12}m^{2}(m-1)\delta_{m+n},\qquad\left[L_{m}^{\pm},L_{n}^{\mp}\right]=0.\label{eq: Virasoro algebra}
\end{equation}
To this end use a more convenient version of the Virasoro for our
contraction purpose, 
\begin{eqnarray*}
\left[J_{m},J_{n}\right] & = & \left(m-n\right)J_{m+n},\\
\left[P_{m},P_{n}\right] & = & \frac{1}{l^{2}}\left(m-n\right)J_{m+n},\\
\left[J_{m},P_{n}\right] & = & (m-n)P_{m+n}+\frac{k}{12}m\left(m^{2}-1\right)\delta_{m+n}.
\end{eqnarray*}
where $k\equiv\frac{c^{+}-c^{-}}{12l}=\frac{3}{G}$. Now we arrrive
at the $\mathfrak{bms}_{3}$ algebra by the simple replacement, $\frac{1}{l}\rightarrow\epsilon,$
(and accordingly $P_{m}\rightarrow\mathcal{P}_{m},J_{m}\rightarrow\mathcal{J}_{m}$)
\begin{eqnarray*}
\left[\mathcal{J}_{m},\mathcal{J}_{n}\right] & = & \left(m-n\right)\mathcal{J}_{m+n},\\
\left[\mathcal{P}_{m},\mathcal{P}_{n}\right] & = & 0,\\
\left[\mathcal{J}_{m},\mathcal{P}_{n}\right] & = & (m-n)\mathcal{P}_{m+n}+\frac{k}{12}m\left(m^{2}-1\right)\delta_{m+n}.
\end{eqnarray*}

\section{Higher Spin Extension}

For the higher spin version, one has to extend the $ISO(2,1)$ algebra
by including a new set of spin-3 generators, $J_{ab},P_{ab}$ \cite{Afshar:2013vka,Gonzalez:2013oaa}. 
The algebra takes the form
\begin{eqnarray*}
\left[J_{ab},J_{cd}\right] & = & -\left(\eta_{a(c}\epsilon_{d)bm}+\eta_{b(c}\epsilon_{d)am}\right)J^{m},\\
\left[J_{ab},P_{cd}\right] & = & -\left(\eta_{a(c}\epsilon_{d)bm}+\eta_{b(c}\epsilon_{d)am}\right)P^{m},\\
\left[P_{ab},P_{cd}\right] & = & 0,\\
\left[J_{a},J_{bc}\right] & = & \epsilon^{m}\,_{a(b}J_{c)m},\\
\left[J_{a},P_{bc}\right] & = & \epsilon^{m}\,_{a(b}P_{c)m},\\
\left[P_{a},J_{bc}\right] & = & \epsilon^{m}\,_{a(b}P_{c)m},\\
\left[P_{a},P_{bc}\right] & = & 0.
\end{eqnarray*}
We will call this the $hsf_{3}$ algebra. The invariant nondegenerate
bilinear product is given by (the only non-vanishing pieces),
\bea
{\rm Tr}(P_{a},J_{b})=\eta_{ab},{\rm Tr}  (P_{ab},J_{ab})=\eta_{ac}\eta_{bd}+\eta_{ad}\eta_{bc}-\frac{2}{3}\eta_{ab}\eta_{cd}. \label{hstrace}
\eea

This algebra can be realized as as Inonu-Wigner contraction of $SL(3, \IR) \times SL(3, \IR)$ algebra analogous to the spin-2 case. In terms of the two copies of the $SL(3)$ generators 
 \bea
&&[T_a,T_b] = \epsilon_{abc} T^c, \\
&&[T_a,T_{bc}] = \epsilon^d_{\ \ a(b} T_{c)d}, \\ 
&&[T_{ab},T_{cd}] = \sigma \left(\eta_{a(c} \epsilon_{d)be}+\eta_{b(c} \epsilon_{d)ae}\right)T^e.
\eea
it can be straightforwardly checked that one can define the $hsf_3$ generators via
\bea
P^a=\left(\begin{array}{cc}
\epsilon \ T^a&0\\
 0 & -\epsilon \ T^a
\end{array}\right), \ &&\ J^a=\left(\begin{array}{cc}
 T^a&0\\
 0 &  T^a
\end{array}\right) \\
P^{ab}=\left(\begin{array}{cc}
\epsilon \ T^{ab}&0\\
 0 & -\epsilon \ T^{ab}
\end{array}\right), \ && \ J^{ab}=\left(\begin{array}{cc}
 T^{ab}&0\\
 0 &  T^{ab}
\end{array}\right).
\eea
This is the Grassmann realization of Inonu-Wigner and our point is that this can be used to interpret a Grassmann valued $SL(3, \IR) \times SL(3, \IR)$ gauge field as an $hsf_3$ (that is, flat space higher spin) gauge field. The Grassmann approach immediately enables us to get to the above result from the generators of \cite{Campoleoni:2010zq}. The traces of the $SL(3,\IR) \times SL(3,\IR)$ are designed so that it reproduces
(\ref{hstrace}). 

As in the spin-2 case, the flat space higher spin gaueg field can be expressed via Grassmann parameter by its natural generalization
\bea
A =\sum_{a=0}^{2}\left(e^{a}+\epsilon\:\omega^{a}\right)T_{a} +\sum_{a,b=0}^{2}\left(e^{ab}+\epsilon\:\omega^{ab}\right)T_{ab}
\eea
Again we expect the actions and the asymptotic symmetries to work out exactly analogously, but we leave the details

\section{Application: Singularity Resolution in the BMS Gauge}

So far what we have done is to merely repeat known results (but from a new and perhaps a simpler and more elegant) point of view. Now we will show that this new technology makes certain computations tractable and show that certain singularity resolution questions become anwerable in this frame work. The reason for this is that constructing a set of explicit matrix generators that satisfy the $hsf_3$ algebra {\em while having the non-degenerate trace form (\ref{hstrace}) } is non-trivial. But one can bypass this problem while having a non-degenerate trace form by working within the Grassmann technology.

Other discussions on singularity resolution in higher spin theories can be found in \cite{Castro:2011fm, Krishnan:2013cra, Krishnan:2013zya, Burrington:2013dda, Krishnan:2013tza}. The basic idea in singularity resolution in this set up is to consider a singular solution of the spin-2 theory, embed it in the higher spin theory, and then to look for gauge transformations that retain the holonomy within the same conjugacy class. If there exists a gauge transformation that gives rise to metric and higher spin fields that are regular while not changing the conjugacy class, we have resolved the singularity.

\subsection{Metric Formulation of the Singular Cosmology}

\begin{figure}
\begin{center}
\includegraphics[
height=0.4\textheight
]{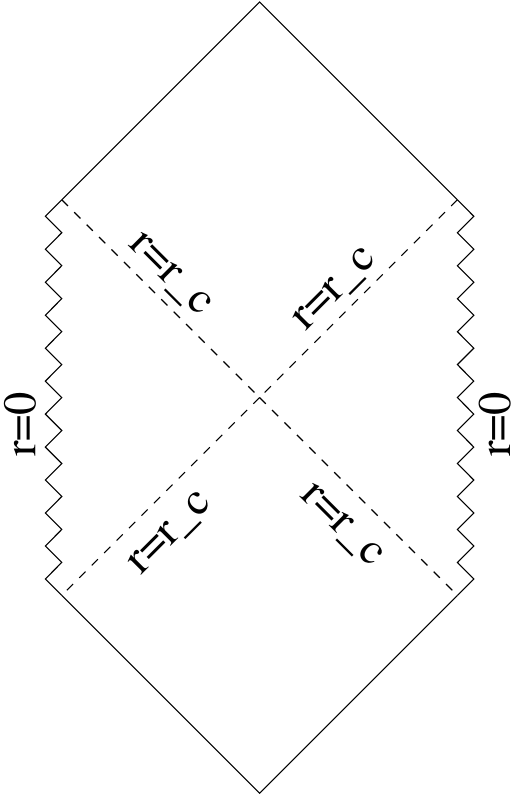}
\caption{Penrose Diagram of the Shifted Boost Orbifold that we Consider. 
}
\label{Penrose}
\end{center}
\end{figure}

We start with the boost-shifted orbifold cosmology \cite{Bagchi:2012xr} which has a sigularity we intend to resolve. To obtain this solution, we can start with the general flat space BMS-gauge solution that we considered previously and specialize to
the case when $\mathcal{M}(\phi)=M$ and $\mathcal{J}(\phi)=J/2$
are constants. As pointed out in \cite{Barnich:2012aw,Barnich:2012xq}
this BMS gauge metric could be thought of as an expression in outgoing
null coordinate, 
\begin{equation}
u=t-\int dr/N^{2}(r),\label{eq: null coordinate}
\end{equation}
and a new angular coordinate,
\begin{equation}
\varphi=\phi-\int dr\; N^{\varphi}/N{}^{2}\label{eq: new angular coordinate}
\end{equation}
to correspond to a Schwarzschild-type metric,
\begin{equation}
ds^{2}=-N^{2}(r)dt^{2}+N^{-2}(r)dr^{2}+r^{2}\left(d\varphi+N^{\varphi}dt\right)^{2},\label{eq: Sch. gauge metric}
\end{equation}
\[
N^{2}(r)=-M+\frac{J^{2}}{4r^{2}}=\frac{M}{r^{2}}\left(r_{C}^{2}-r^{2}\right),N^{\varphi}=\frac{J}{2r^{2}}.
\]
Note that $\varphi=\phi-\int dr\; N^{\varphi}/N{}^{2}$ and hence
does $not$ parametrize a compact direction. However one can identify
$\varphi\sim\varphi+2\pi$ and construct quotient spaces \cite{Cornalba:2002fi,Cornalba:2003kd,Barnich:2012aw}
with cosmological (Cauchy) horizons at $r=r_{C}$. However these spaces
contain pathological regions with closed time-like curves, $r<0$
and such regions are excised. $r=0$ thus becomes a causal structure
singularity. These have been dubbed \emph{shifted boost orbifolds}
\cite{Cornalba:2002fi,Cornalba:2003kd, Berkooz:2002je}\footnote{The reason for the name is the fact that the metric can be understood as an orbifold of flat space under shifts and boosts, but we will not need that connection, so we will not elaborate on it.} when they were discovered
and discussed in the context of string theory. However, we shall refer
to these as flat quotient cosmologies. A Penrose diagram of the flat
quotient cosmology is provided in Fig. \ref{Penrose}. 


\subsection{Gauge Theory Formulation of the Singular Cosmology}
 
For completeness, we present 
the expressions for the set of triads and the dual spin connection
for the flat quotient cosmology in Schwarzschild gauge (\ref{eq: Sch. gauge metric}):
\begin{equation}
e^{0}=N(r)dt,\qquad e^{1}=-N^{-1}(r)dr,\qquad e^{2}=rN^{\varphi}(r)\; dt+rd\varphi,\label{eq: triads}
\end{equation}
\begin{equation}
\omega^{0}=N(r)d\varphi,\qquad\omega^{1}=\frac{N^{\varphi}(r)}{N(r)}\; dr,\qquad\omega^{2}=r\; N^{\varphi}(r)d\varphi.\label{eq: dual spin connections}
\end{equation}
The triad and the spin connection for the general BMS gauge solution has been written down before. We will need the full gauge connection that we wrote down in (\ref{eq: BMS gauge full connection}). 

It turns out that one can express this full connection, $A$ in terms of a \emph{primitive}
connection, $a$ which is stripped-off of any $r$-dependence,
\begin{eqnarray}
a(u,\phi) & = & -\frac{1}{\sqrt{2}}\left[\left(\frac{\mathcal{M}}{2}-1\right)\: du+\left(\mathcal{J}+\frac{u}{2}\mathcal{M}'\right)\: d\phi\right]P_{0}\nonumber \\
 &  & \qquad\qquad\qquad-\frac{1}{\sqrt{2}}\left[\left(\frac{\mathcal{M}}{2}+1\right)\: du+\left(\mathcal{J}+\frac{u}{2}\mathcal{M}'\right)\: d\phi\right]P_{1}\nonumber \\
 &  & \hspace{-0.5in}\qquad\qquad\qquad\qquad\qquad\qquad-\frac{1}{\sqrt{2}}\left(\frac{\mathcal{M}}{2}-1\right)\: d\phi\: J_{0}-\frac{1}{\sqrt{2}}\left(\frac{\mathcal{M}}{2}+1\right)\: d\phi\: J_{1},\label{eq:primitive connection}
\end{eqnarray}
using a radial gauge transformation, $b(r)=\exp\left(\frac{rP_{0}+rP_{1}}{\sqrt{2}}\right)$,
\[
A=b^{-1}\: a\: b+b^{-1}\partial_{r}b.
\]
This will be useful to us in resolving the singularity.

\subsection{Holonomy of the flat cosmology}

For the flat quotients, $\mathcal{M}(\phi)=M$ and $2\mathcal{J}(\phi)=J$
are constants, and the Wilson loop operator along a constant $u$,
$\phi$-circle around $r=0$ is,
\[
W=\exp\left(\int_{0}^{2\pi}d\phi\; A\right)=\exp\left(b^{-1}2\pi a_{\phi}b\right)=b^{-1}\exp\left(2\pi a_{\phi}\right)b,
\]
where,
\[
a_{\phi}=-\frac{1}{\sqrt{2}}\left[\frac{J}{2}\left(P_{0}+P_{1}\right)+\left(\frac{M}{2}-1\right)\: J_{0}+\left(\frac{M}{2}+1\right)\: J_{1}\right].
\]
Under a trivial gauge transformation, $U$,
\[
W=UWU^{-1}=\exp^{UwU^{-1}}
\]
where, $w\equiv2\pi a_{\phi}$. 

\subsection{$ISO(2,1)$ Solution as a Grassmann Valued $SO(2,2)$ Solution\label{sec: From Grassmann SL(2,R) to ISO(2,1)}}

We haven't introduced explicit matrices for the $P^a$ and $J^a$, but we do not need to. This is because the same holonomy information can be captured equivalently in the Grassmann language.  To this end, we first note that the full connection can also be written as the Grassmann valued connection (\ref{eq: BMS gauge full connection-1}). 

This connection can also be written in terms of a primitive connection. The radial dependence is contained in the term,
\bea
A_{r}=\epsilon\frac{T_{0}+T_{1}}{\sqrt{2}}dr.
\eea
We can try to gauge away $r$-dependence and construct a primitive
connection by gauge transforming with 
\bea
U=\exp(\epsilon T_{+}r), 
\eea
with $ T_{\pm}=\frac{T_{0}\pm T_{1}}{\sqrt{2}}$,
so that $A=U^{-1}aU+U^{-1}dU$.
This $r$-independent primitive connection is
 \begin{equation}
a(u,\phi)= \left[ \epsilon T_{-} - \epsilon \frac{M}{2} T_{+} \right]du + \left[ -\left( \frac{M}{2} + \epsilon \left(J + \frac{u}{2}M' \right) \right)T_{+} + T_{-} \right]d\phi
\label{eq:Primitive connection in SL(2) sector}
\end{equation}
We are interested in computing the holonomy matrix of this connection $a$ along a $\phi$-circle of constant $u,r$ in the special case when $M$, $J$ are constant. The eigenvalues of the holonomy are given by 
\begin{equation}
w=2\pi a_{\phi} = \{0,\ -2\pi \sqrt{M} -\pi \frac{J}{\sqrt{M}}\epsilon,\ 2\pi \sqrt{M} +\pi \frac{J}{\sqrt{M}}\epsilon \} \label{eq:grassmann valued holonomy matrix}
\end{equation}

For more generic cases determining the eigenvalues is hard, so instead we use the characteristic polynomial theorem for $3\times3$ square matrices. For a square matrix $M$, the characteristic equation is
\begin{equation}
M^{3}= \mbox{M} \mathbb{I}_{3}+\frac{1}{2}\left(\mbox{tr}(M^{2})-(\mbox{tr}(M))^{2}\right)M+\mbox{tr}(M)\: M^{2}.\label{eq: Characteristic Polynomial Expansion}
\end{equation}
The eigen-values of two holonomy matrices are identical, iff the coefficients of their characteristic polynomials agree. It is easy to check that this theorem is valid even when the matrix has Grassmann valued matrix elements. 

For the flat cosmology, and $M=w$, the left and right holonomy matrices give rise to
\begin{eqnarray}
\mbox{Det}w & = & 0,\nonumber \\
\mbox{tr}\left(w\right) & = & 0,\nonumber \\
\mbox{tr}\left(w_{\pm}^{2}\right) & = & 8\pi^{2} \left(M + 2\epsilon\: J\right) \label{eq: Holonomy conditions}
\end{eqnarray}
Of course, since $\exp(w_{\pm})\in SL(3)$, $\mbox{tr}(w_{\pm})=0$
is automatically ensured.%

\subsection{Singularity Resolution}

We extend the $SL(2,R)$ connection (\ref{eq:Primitive connection in SL(2) sector}) by adding Grassmann valued $SL(3)$ generators,
\begin{equation}
a' = a + \sum_{a,b=0}^{2}\left(c_{ab}+\epsilon\: d_{ab}\right)T_{ab}.\label{eq: SL(3,R) X SL(3,R) primitive}
\end{equation}
After gauge transforming to include the radial dependence we will have a form,
\bea
 A'=A + \sum_{a,b=0}^{2}\left(e_{ab}+\epsilon\:\omega_{ab}\right)T_{ab} 
\eea
This is a Grassmann valued $SL(3,R)\times SL(3,R)$ connection and equivalently a connection in the higher spin theory in asymptotically flat space. The metric and the higher spin fields can be obtained from the gauge field by identifying the triad (and its higher spin version)\cite{Gonzalez:2013oaa}.
The correction to metric takes the explicit form
\bea
ds^{2}=\left(\eta_{ab}e^{a}\,_{\mu}e^{b}\,_{\nu}+2\eta_{ac}\eta_{bd}e^{ab}{}_{\mu}e^{cd}\,_{\nu}\right)dx^{\mu}dx^{\nu}.\label{eq:metric components after higher spin corrections}
\eea

Actually, instead of using the generators, $T_{ab}$ which do not constitute a linearly independent set, we will use the set, $W_{a}$ \cite{Castro:2011fm}
\bea
a'=a+\sum_{a=-2}^{2}\left(c^{a}+\epsilon\: d^{a}\right)W_{a}  \\ A'=A+\sum_{a=-2}^{2}\left(C^{a}+\epsilon\: D^{a}\right)W_{a} 
\eea
We may look at the simplest case of singularity resolution where we only turn on $W$ generators in $a_{\phi}$ component of the primitive connection.
\begin{equation}
a'_{\phi} = a_{\phi} + \sum_{a=-2}^{2} (c_a + \epsilon d_a)W_a
\end{equation}
Next we need to satisfy the equations of motion i.e flatness of the connection, 
\bea 
da'+a'\wedge a'=0 
\eea
\begin{itemize}
\item Demanding $c_{r},d_{r}=0$ i.e., no radial components, flatness implies the coefficients $c^{a}$ and $d^{a}$ are independent of $r$. This should not be surprising as this is still in ``radial'' gauge or a primitive connection, where radial dependence has been gauged away just like in the $SL(2)$ sector, 
\bea
c_{\mu}^{a}=c_{\mu}^{a}(u,\phi) \qquad d_{\mu}^{a}=d_{\mu}^{a}(u,\phi) 
\eea
But the surprising fact that higher spin contribution to the metric is \emph{$r$ independent} as evident from Eq. (\ref{eq:metric components after higher spin corrections}).

\item We futhermore assume all coefficients in the primitive connection to be constants. This is justified because we care to find {\em some} resolution, not the most general resolution of the singularity. The equation of motion for these coefficients are then given by
\begin{equation}
[a_u,a_{\phi}] = 0
\end{equation}
This gives us following conditions
\begin{eqnarray}
c_1 = 0, \ \ c_{-1} =0, \ \ c_0 + Mc_{-2} = 0, \ \ M c_0 + 4c_2 = 0 . 
\end{eqnarray}
Coefficients $d_1$ and $d_{-1}$ are not determined by any equation and can be freely choosen to be zero.
\end{itemize}
Next, we impose the holonomy constraints. As before we want the eigenvalues of
$w=2\pi a'_{\phi}$ to be
same as that of Eq. (\ref{eq: Holonomy conditions}).

\begin{enumerate}
\item The trace condition gives,
\begin{eqnarray}\nonumber
\frac{8c_0^2}{3} + 32c_2 c_{-2} =  0, \hspace{0.5in} \\ \nonumber
\frac{16c_0 d_0}{3} + 32 c_{-2} d_2 + 32 c_2 d_{-2} = 0. \nonumber
\end{eqnarray}
\item Determinant condition gives,
\begin{eqnarray}\nonumber
-\frac{16 c_0^3}{27} + 4c_2 + \frac{64c_0 c_2 c_{-2}}{3} - \frac{2c_0 M}{3} + c_{-2}M^2 = 0, \hspace{1in}\\ \nonumber
-\frac{16c_0^2 d_0}{9} + \frac{64 c_2 c_{-2} d_0}{3} + 4d_2 + \frac{64c_0 c_{-2} d_{-2}}{3} - \frac{4c_0 J}{3} - \frac{2d_0 M}{3} + 4c_{-2}JM + d_{-2}M^2=0. \nonumber
\end{eqnarray}
\end{enumerate}
These equations can be consistently solved for various coefficients $c_a$ and $d_a$. Here we list one particular solution which helps in singularity resolution.
\begin{eqnarray}
c_2 = 0, \ \ c_{-2} = 0, \ \ c_0 = 0,
\end{eqnarray}
together with coefficients $d_0$, $d_{2}$ and $d_{-2}$ which are now constrained to obey following relation
\bea
4d_2 - \frac{2d_0 M}{3} + d_{-2}M^2 = 0
\eea
Transforming back to full $r$-dependent gauge, we obtain the metric to be
\begin{equation}
ds^{2} = \mathcal{M}du^{2}-2dudr+2\mathcal{J}dud\phi + \left[ \frac{12d_0^2}{9} + 16d_2 d_{-2} + r^2 \right] d\phi^{2}
\end{equation}
The fact that the collapsing $\phi$-cycle is now stabilized at finite radius is evident from the metric. More concretely, it can also be seen explicitely from the form of Ricci scalar
\begin{equation}
R = \frac{24\left(d_0^2 + 12d_2 d_{-2} \right)M}{\left( 4d_0^2 + 48d_2 d_{-2} + 3r^2  \right)}
\end{equation}
which is a non-constant, but everywhere non-singular function of $r$. It can also be checked that the higher spin fields that result from the gauge transformation are also regular everywhere, even though we will not present the details. 

In any event, singularity resolution was only illustrative for our purposes here: our goal was to demonstrate that the Grassmann approach can be a useful technical tool and not merely a curiosity.


\section*{Acknowledgments}

CK thanks Glenn Barnich for discussions (of yore) on the BMS algebra, Rudranil Basu for helpful clarifications on Chern-Simons gauge theories, and Arjun Bagchi for raising questions on the interpretation of $G$ which improved the presentation of the final draft. The research of SR is supported by Department of Science and Technology (DST), Govt.$\hspace{0.05in}$of India research grant under scheme DSTO/1100 (ACAQFT). 
\appendix


\bibliographystyle{brownphys.bst}
\bibliography{ads-to-flat.bib}

\end{document}